\pdfoutput=1 %
\documentclass[sigconf, nonacm]{acmart}

\usepackage{hyperref}
\usepackage{cleveref}   %
\usepackage{pgfplots}\pgfplotsset{compat=1.18}
\usepgfplotslibrary{groupplots}
\usepackage{siunitx}
\usepackage{tikz}
\usetikzlibrary{arrows.meta, backgrounds, fit, positioning, calc, shapes}
\tikzset{ %
	extends/.style={->, >={Triangle[open, width=0.25cm, length=0.25cm]}}
}
\usepackage{tabularx}
\usepackage{listings}
\usepackage{xcolor}
\usepackage{colortbl}
\usepackage{subcaption}
\usepackage{amssymb}
\usepackage{pifont}
\usepackage{listings}
\usepackage{booktabs}

\lstdefinelanguage{Cypher}{
  keywords={MATCH, RETURN, WHERE, CREATE, DELETE, SET, MERGE, OPTIONAL, WITH, AS, AND, OR, NOT, LIMIT, ORDER, BY},
  sensitive=true,
  morecomment=[l]{//},
  morestring=[b]"
}

\AtBeginDocument{%
  }

\setcopyright{cc}
\copyrightyear{2026}
\acmYear{2026}
\acmDOI{10.1145/3841645.3843314}
\acmConference[SIGSPATIAL '26]{The 34th ACM International Conference on Advances in Geographic Information Systems}{November 03--06, 2026}{Riverside, CA, USA}

\acmBooktitle{The 34th ACM International Conference on Advances in Geographic Information Systems (SIGSPATIAL '26), November 03--06, 2026, Riverside, CA, USA}
\acmISBN{979-8-4007-2950-8/2026/11}

\newcommand{\mytoolname}{\textit{pykci}} %
\newcommand{\mytoolsound}{\textit{pixie}}

\setcctype{by}

\begin{document}

\sloppy
\title{pykci: A Compact Urban Knowledge Graph for Semantic and Spatial Queries using LLMs}

\author{Huynh Duc An Son Nguyen}
\email{son.nguyen@hcu-hamburg.de}
\orcid{https://orcid.org/0000-0001-8711-1587}
\affiliation{%
  \institution{HafenCity University Hamburg, \\ Computational Methods Lab}
  \streetaddress{Henning-Voscherau-Platz 1}
  \city{Hamburg}
  \state{}
  \country{Germany}
  \postcode{20457}
}

\author{Lukas Arzoumanidis}
\email{lukas.arzumanidis@hcu-hamburg.de}
\orcid{https://orcid.org/0000-0001-6668-1695}
\affiliation{%
  \institution{HafenCity University Hamburg, \\ Computational Methods Lab}
  \streetaddress{Henning-Voscherau-Platz 1}
  \city{Hamburg}
  \state{}
  \country{Germany}
  \postcode{20457}
}

\author{Youness Dehbi}
\email{youness.dehbi@hcu-hamburg.de}
\orcid{https://orcid.org/0000-0003-0133-4099}
\affiliation{%
  \institution{HafenCity University Hamburg, \\ Computational Methods Lab}
  \streetaddress{Henning-Voscherau-Platz 1}
  \city{Hamburg}
  \state{}
  \country{Germany}
  \postcode{20457}
}

\renewcommand{\shortauthors}{Nguyen et al.}

\begin{abstract}
CityGML, the OGC standard for the modeling, storage, and exchange of semantic 3D city models, describes urban objects with detailed semantics, geometry, and topology. Yet this richness is difficult to query directly: CityGML's XML encoding is designed for exchange rather than analysis, and relational mappings expose it through schemas that demand expert knowledge. We present \textit{\mytoolname} (Python Knowledge Graph for Cities), an open-source system that transforms CityGML 2.0 datasets into a compact urban knowledge graph in Neo4j and makes it queryable in plain natural language. The graph schema covers all thematic feature modules of CityGML 2.0 across all levels of detail and is spatially indexed with an R-tree for efficient geometric retrieval. A complete end-to-end Python pipeline ingests CityGML datasets into the knowledge graph, exports them to OGC 3D Tiles for interactive visualization, and supports lossless round-trip export of all modeled content back to CityGML. For querying, the graph is paired with a large language model through a model-agnostic text-to-Cypher mechanism: the graph schema is supplied as context, and the model translates natural-language questions into Cypher queries executed directly against the graph. We evaluate both a locally running open-weight model, which keeps sensitive city data on-premise, and a state-of-the-art commercial model for the most demanding spatial and semantic queries. Answers are thereby grounded in exact city data rather than the model's parametric memory, reducing hallucination and providing auditable provenance for every response. We demonstrate the system on open-government CityGML LoD2 datasets from Hamburg, Germany, including complex semantic and spatial queries such as identifying roof surfaces suitable for greening. \textit{\mytoolname} enables urban planners, GIS practitioners, and citizens to interact with semantic 3D city models without expertise in query languages and database schemas. The project is publicly available at \url{https://github.com/hcu-cml/pykci}.
\end{abstract}

\ccsdesc[500]{Information systems~Graph-based database models}
\ccsdesc[500]{Information systems~Geographic information systems}
\ccsdesc[300]{Computing methodologies~Natural language generation}

\keywords{Knowledge Graph, Database, Semantic, Spatial, LLM, CityGML}
\begin{teaserfigure}
  \includegraphics[width=\textwidth]{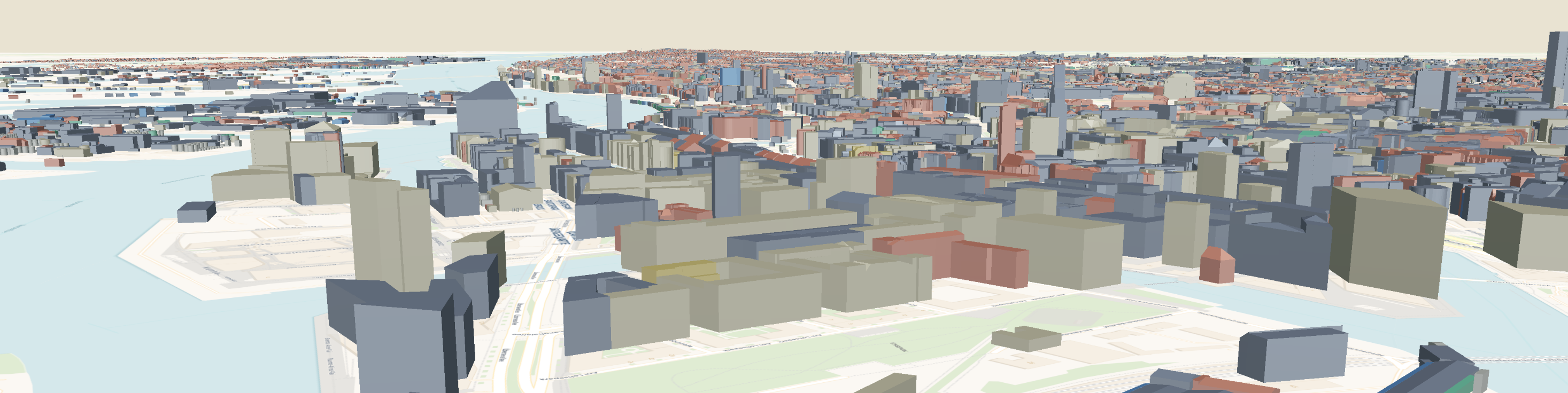} %
  \caption{OGC 3D Tiles visualization of CityGML LoD2 dataset of Hamburg, automatically exported from its urban knowledge graph constructed by~\mytoolname. Buildings are colored based on their roof materials stored in the graph. Base map data from OpenStreetMap without elevation.}
  \Description{A 3D rendering of multiple buildings in Hamburg, each colored based on their roof materials. Base map data from OpenStreetMap without elevation.}
  \label{fig:teaser}
\end{teaserfigure}

\maketitle

\section{Introduction}\label{sec:introduction}

More than 4.2 billion people, over \SI{55}{\percent} of the global population,
live in cities today, and the United Nations projects this proportion will
reach \SI{68}{\percent} by 2050~\cite{un_urban_2018}. Supporting this growth
imposes increasing demands on infrastructure, urban planning, and data
management. In Germany alone, the national 3D Building Model at Level of Detail
2 (LoD2-DE) already comprises 58 million buildings~\cite{adv_buildings_2025},
each represented with detailed semantics, geometry, and topology: typed boundary surfaces,
precise three-dimensional geometry, and thematic attributes such as function,
roof form, and measured height. Such data increasingly forms the backbone of
\emph{urban digital twins} (UDTs), virtual replicas of the physical city used
for planning, simulation, and decision-making~\cite{ketzler_udtreview_2020,lei_udt_survey_2023}.
At the core of any UDT lies a queryable digital knowledge base, and how easily
its encoded knowledge can be accessed by various stakeholders determines how useful the twin is in
practice.

The City Geography Markup Language (CityGML)~\cite{groeger_citygml2_2012}, an
Open Geospatial Consortium (OGC) standard for modeling and exchanging urban
objects and landscapes, is the de facto format for semantic 3D city models, but
its XML encoding is designed for exchange rather than analysis. The most widely
used backend, the 3D City Database (3DCityDB)~\cite{yao_3dcitydb_2018}, maps
CityGML to a relational schema spanning many interrelated tables; retrieving
semantically meaningful relationships, such as identifying buildings in a
district suitable for roof greening given constraints on roof slope, type, and
material, requires complex multi-join queries that scale poorly and obscure the
data's inherent structure. That structure is fundamentally graph-like: a
building decomposes into boundary surfaces, then polygons, then points, and
CityGML reuses shared geometry, such as a wall surface referenced by two
adjacent rooms, through explicit references. Such deeply connected data is more
naturally represented in a graph than in flat relations~\cite{nguyen_phdthesis_2024}.

Graph databases store entities as nodes and relationships as first-class edges,
enabling efficient traversal of connected data~\cite{robinson_graphdb_2015},
and knowledge graphs (KGs) add a semantic layer of typed entities and
relations~\cite{hogan_kgs_2021,pan_llmskgs_2024}. KGs are increasingly adopted
as the integrating knowledge base inside UDTs~\cite{akroyd_kgudt_2021,ramonell_kgudt_2023,saifwajid_kgudt_2024}.
We represent CityGML as a labeled property graph (LPG) in Neo4j rather than as
Resource Description Framework (RDF) triples, a choice we motivate in \Cref{sec:related_work}; and while the
closest prior system, 3DCityKG~\cite{nguyen_phdthesis_2024}, also maps CityGML
to a graph, it mirrors the verbose CityGML hierarchy, whereas we adopt a compact
schema with shorter traversal paths.

A KG does not by itself lower the query barrier: users must still know the
schema and a query language such as Cypher. Large language models (LLMs) promise
a natural-language interface, but applying them directly to city models fails on
two counts. A single city's CityGML describes tens to hundreds of thousands of buildings,
far exceeding practical context limits and prohibitive in token cost; and
correct interpretation demands domain-specific knowledge, such as the
distinction between a \textit{GroundSurface} and a \textit{RoofSurface} or the
meaning of code-based function values, that general-purpose models lack.
Embedding a city into the model itself, as in CityGPT~\cite{feng_citygpt_2025},
conflates the query interface with the data store and does not generalize across
cities or updates. We instead keep the knowledge base external: the LLM receives
the graph schema as context and generates Cypher that executes directly against
the graph. This keeps context compact, lets the data be updated without
retraining, and grounds every answer in exact, auditable graph data rather than
the model's parametric memory.

We present \textit{\mytoolname} (Python Knowledge Graph for Cities, pronounced
\mytoolsound), a compact urban KG for CityGML 2.0, implemented as an LPG in
Neo4j. The graph schema covers all thematic feature modules of CityGML 2.0,
including buildings, bridges, tunnels, transportation, and vegetation, across
all five levels of detail (LoD0--4), and incorporates an R-tree spatial index
for efficient geometric querying. We provide a complete end-to-end pipeline in
Python that ingests CityGML datasets into the KG, exports the result to OGC 3D
Tiles for interactive 3D visualization, and supports round-trip export back to
CityGML for editing workflows. For complex querying, the KG is paired with an
LLM through a model-agnostic text-to-Cypher interface: a locally running
open-weight model serves routine queries entirely on-premise, while a commercial
model handles the most demanding reasoning when higher accuracy is required 
(rubric in~\Cref{tab:query-detail}). 
The
generated queries run directly against the graph, enabling urban planners and
GIS practitioners to interact with 3D city data without query-language
expertise. \mytoolname\ is demonstrated on open-government CityGML LoD2 datasets
from Hamburg, Germany, is released as open-source software, and, within the
broader vision of UDTs, serves as the queryable digital knowledge base that
makes a city's semantic 3D model accessible to both software components and
non-expert users. Our main contributions are: (1) a compact, semantically and
spatially indexed urban KG schema for CityGML 2.0; (2) a fully open-source,
end-to-end ingestion, visualization, and export pipeline in Python; and (3) a
natural-language querying interface for complex semantic and spatial queries
over large-scale 3D city models via LLM.

\begin{figure*}[t]
  \centering
  \definecolor{wfInB}{RGB}{0,105,92}    \definecolor{wfInF}{RGB}{224,242,241} %
  \definecolor{wfKgB}{RGB}{239,108,0}   \definecolor{wfKgF}{RGB}{255,243,224} %
  \definecolor{wfVzB}{RGB}{46,125,50}   \definecolor{wfVzF}{RGB}{232,245,233} %
  \definecolor{wfLmB}{RGB}{106,27,154}  \definecolor{wfLmF}{RGB}{243,229,245} %
  \definecolor{wfUsB}{RGB}{69,90,100}   \definecolor{wfUsF}{RGB}{236,239,241} %
  \definecolor{wfTwB}{RGB}{38,50,56}    \definecolor{wfTwF}{RGB}{247,249,251} %
  \includegraphics{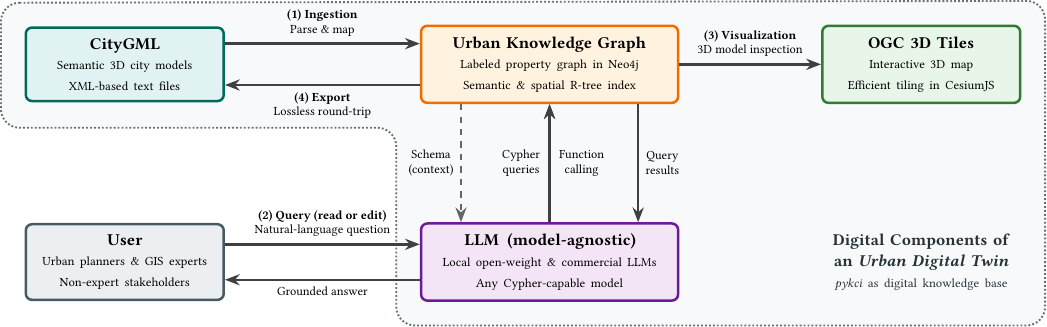}
  \caption{Overview of \textit{\mytoolname} as the digital knowledge base of an
  \textit{urban digital twin}. A CityGML dataset is ingested into an urban
  knowledge graph in Neo4j (1) and queried in natural language (2): the LLM
  receives the graph schema as context and accesses the database either by
  calling predefined functions or by generating Cypher queries directly. The
  graph can be exported to OGC 3D Tiles for visualization (3) and back to
  CityGML (4). The querying LLM is model-agnostic: a local open-weight model
  for routine, on-premise queries, and a commercial model for the most
  demanding ones.}
  \Description{Overview diagram of \textit{\mytoolname} as the digital knowledge base of an urban digital twin.}
  \label{fig:workflow}
\end{figure*}

\Cref{fig:teaser} visualizes the resulting urban KG for Hamburg using  OGC 3D Tiles, where the LoD2 3D building models are colored based on their roof materials stored in the KG. 
\Cref{fig:workflow} gives an
overview of the workflow, positioning \mytoolname\ as the digital knowledge base
of a UDT.
\Cref{sec:related_work} reviews related work; 
\Cref{sec:graph_schema} presents the \mytoolname\ schema and mapping rules; %
\Cref{sec:evaluation} evaluates the proposed mapping method based on well-known datasets; 
\Cref{sec:benchmarking} assesses the performance and efficiency of \mytoolname\ against related tools;
\Cref{sec:usecases_enrichment} demonstrates real-world LLM-based querying use cases; and 
\Cref{sec:conclusion} concludes.

\section{Related Work}\label{sec:related_work}

This section briefly reviews prior work related to \mytoolname: semantic 3D city models,  CityGML data model, relational and graph representations of CityGML, urban KGs and UDTs, and LLMs in combination with spatially-enhanced KGs.

\subsection{Semantic 3D City Models and CityGML}
CityGML~\cite{groeger_citygml2_2012} is the OGC standard for representing,
storing, and exchanging semantic 3D city models, and one of the key data models
for urban digital twins. Unlike conventional 3D city models, which capture only
graphical or geometrical representation, CityGML jointly encodes the semantic,
topological, geometrical, and appearance properties of city objects in a single
model. It is an application schema of GML~3.1.1~\cite{cox_gml_2004}, the OGC's
XML grammar for geographic features defined in line with ISO/TC~211. 
Released in successive versions (1.0 in 2008, 2.0 in 2012, and 3.0 in 2021~\cite{kolbe_citygml3_2021}), 
CityGML has become the most widely used exchange format for semantic 3D city models. 
It has been adopted in more than 20 countries~\cite{wysocki_citygml_2022}, 
including Japan, Australia, Germany, the United Kingdom, and the United States.
This paper focuses on CityGML~2.0, which remains the most widely adopted version and 
accounts for the majority of open-government datasets, including the dataset used in this study. 
Unless stated otherwise, the term ``CityGML'' hereafter refers to version~2.0.
Alternative encodings exist: CityJSON~\cite{ledoux_cityjson_2019} provides a
compact JSON serialization of a subset of CityGML's data model, while
OpenStreetMap (OSM) offers crowd-sourced but largely LoD1 building data and street networks.

CityGML~\cite{groeger_citygml2_2012} represents a city as a collection of
\textit{city objects} organized into thematic modules 
(\textit{Appearance}, \textit{Bridge},
\textit{Building}, \textit{City\-Furniture}, \textit{City\-Object\-Group},
\textit{Generics}, \textit{Land\-Use}, \textit{Relief}, \textit{Transportation},
\textit{Tunnel}, \textit{Vegetation}, \textit{Water\-Body}, and now-deprecated
\textit{Textured\-Surface}) 
and across five levels of detail (LoD0--4). 
A building is bounded by
typed \emph{boundary surfaces}, namely \textit{Ground\-Surface},
\textit{Wall\-Surface}, and \textit{Roof\-Surface}, each described by one or more
polygons. A polygon is defined by an exterior ring 
and optional interior rings, and rings are ordered lists of points. This yields
a strict containment hierarchy
that at first appears tree-like.
In practice, however, geometry can be shared
between features through references (XLinks), so that a single surface instance
is defined once and reused by multiple objects. The data model is therefore a
graph rather than a tree~\cite{nguyen_changedetection_2017}. 

To enable this semantic and structural richness, CityGML employs a complex, multi-level class hierarchy in which elements are defined at different levels of abstraction~\cite{groeger_citygml2_2012}. It makes extensive use of object-oriented principles such as inheritance, encapsulation, and polymorphism. 
In contrast, graph databases such as Neo4j, like many other database systems, are instance-based~\cite{robinson_graphdb_2015}, meaning they primarily store instances of classes rather than explicit class definitions. 
Another challenge arises from the permissive nature of the GML and CityGML encoding standard, which allows multiple syntactic representations for the same object~\cite{nguyen_phdthesis_2024}. For example, a surface can be represented either as a single polygon or as a collection of smaller adjacent polygons that together form the same geometry. This syntactic ambiguity complicates the conversion between CityGML and graph-based data models, particularly when aiming to produce a compact graph representation, such as \mytoolname.

\subsection{Database Representations of CityGML}
Representing CityGML in a database poses several challenges.
Production datasets are large, reaching tens of millions of
buildings~\cite{adv_buildings_2025}; the model has a deep class hierarchy;
geometry is frequently shared between features through references (XLinks); and
the underlying encoding schemas allow several syntactic ways to define
the same object. Together, these properties render traditional relational or document
storage either lossy or inefficient for complex analytical querying.

Relational databases remain the most widely used backend for semantic 3D city models. 3DCityDB~\cite{yao_3dcitydb_2018}
maps CityGML to many interrelated tables and is well suited to storage and
exchange, but recovering semantic relationships requires multi-join queries that
obscure the data's connected structure. To better preserve this structure,
object-oriented and graph-based mappings have been proposed.

Two graph paradigms dominate. In the semantic-web tradition, CityGML is mapped
to RDF triples with ontologies in the Web Ontology Language (OWL) and GeoSPARQL
as the spatial query interface~\cite{chadzynski_sem3dcitydb_2021,ding_osmgraph_2025},
enabling linked-data integration and ontology reasoning. However, standard RDF cannot attach properties to
edges~\cite{angles_edgeprops_2017}. Encoding the attributes of a relationship
instead requires reification, which inflates triple counts and complicates
querying. Such relationship attributes are common in 3D city models, for example, the ordering 
of boundary surfaces within a building and the XLinks present in the original CityGML dataset.
\mytoolname\ preserves these natively as edge properties for reconstruction of CityGML on export.
The RDF-star extension~\cite{hartig_rdfstar_2017} allows triple annotation
by allowing statements about statements, but its support across
triplestores and GeoSPARQL implementations remains uneven. It also does not
alter RDF's separation of spatial evaluation from graph traversal: 
GeoSPARQL evaluates spatial predicates over geometry literals such as
Well-known Text (WKT) via filter functions~\cite{car_geosparql_2024} rather than
as native graph traversal primitives.

In the LPG tradition, 3DCityKG~\cite{nguyen_phdthesis_2024} maps CityGML to a
Neo4j graph with high-performance, multi-threaded ingestion. Its graph mirrors
the CityGML class hierarchy, which enables lossless round-trip export and
intuitive navigation for domain experts, but places content nodes many hops
below their top-level features, increasing traversal depth and the context cost
of querying. Originally designed for automatic change detection and interpretation 
of large semantic 3D city models within the tool citymodel-compare~\cite{nguyen_patterns_2023,nguyen_phdthesis_2024}, 
3DCityKG features an in-memory spatial index for efficient online graph matching, 
but the persistent graph cannot leverage it in its current version~\cite{liu_kcitychatbot_2025}.
The appropriate
representation thus depends on the use case: exchange and archival favor
3DCityDB, ontology integration favors RDF, and analytical traversal, spatial
querying, and compact LLM context favor a purpose-built LPG. \mytoolname\ follows
the last direction with a deliberately compact schema.

\subsection{Neo4j, Cypher, and Spatial Indexing}

Neo4j is a graph database that implements the LPG
model~\cite{robinson_graphdb_2015}. It stores nodes and relationships with
native pointers (index-free adjacency), allowing constant-time traversal 
of relationships independent of dataset size. Graphs are queried in
Cypher, a declarative, pattern-based query language in which queries 
specify subgraph patterns to match. For example, the following query 
retrieves all buildings together with their roof surfaces:
\begin{lstlisting}[language=Cypher]
MATCH (b:Building)-[:HAS_BOUNDARY]->(s:RoofSurface)
RETURN b, s
\end{lstlisting}
In addition to the built-in labeled-base semantic and point indexes, 
\mytoolname\ uses the Neo4j Spatial extension~\cite{neo4j_spatial_2025}, which
maintains an R-tree index over feature 2D bounding boxes and exposes spatial
predicates such as bounding-box, intersection, and within-distance search. The
combination of pattern-based traversal and spatial indexing allows a single
query to express both semantic conditions (for example, roof type) and spatial
conditions (for example, proximity to a location) over the same graph.

\subsection{Urban Knowledge Graphs and Digital Twins}
The term ``knowledge graph'' has no single agreed
definition~\cite{ehrlinger_kgdef_2016,bonatti_kgdef_2018,bergman_kgdef_2019,hogan_kgs_2021}. 
We adopt the definition of Hogan et al.~\cite{hogan_kgs_2021}: a knowledge graph
(KG) is ``a graph of data intended to accumulate and convey knowledge of the
real world, whose nodes represent entities of interest and whose edges represent
relations between these entities.'' A KG thus adds a semantic layer to a graph
by fixing the meaning of node and edge types through a schema or
ontology~\cite{pan_llmskgs_2024}. 
Surveys of the field~\cite{wang_urbankgreview_2024,pan_llmskgs_2024} distinguish
general or world KGs, such as DBpedia~\cite{lehmann_dbpedia_2015},
Wikidata~\cite{vrandecic_wikidata_2014}, and YAGO~\cite{hoffart_yago2_2011}, from
domain-specific KGs.
Urban KGs apply the concept at city scale: UrbanKG~\cite{liu_urbankg_2023} fuses
regions, points of interest, and mobility data to support traffic prediction and
site selection, while UrbanKGent~\cite{ning_urbankgent_2024} uses LLMs to assist
urban KG construction. Reviews note that smart-city domain KGs remain at an early
stage of development~\cite{wang_urbankgreview_2024}. KGs are increasingly used as
the integrating knowledge base of UDTs~\cite{lei_udt_survey_2023,akroyd_kgudt_2021,ramonell_kgudt_2023,saifwajid_kgudt_2024},
linking heterogeneous data into a queryable structure. Most existing urban KGs,
however, emphasize point-of-interest and mobility data rather than the detailed
geometric and semantic content of CityGML LoD2 models, which is the focus of our
work.

\subsection{LLMs and Spatial Knowledge Graphs}
LLMs and KGs are increasingly combined. \citet{pan_llmskgs_2024} survey
this synergy, observing that KGs supply external, updatable, and interpretable
knowledge while LLMs contribute language understanding, and that KGs are by
nature difficult to construct and evolve. Retrieval-augmented generation
(RAG)~\cite{lewis_rag_2020} grounds LLM outputs in retrieved external data, and
GraphRAG~\cite{edge_graphrag_2025} extends this to graphs built from
unstructured text. Our setting differs fundamentally: the graph is an
authoritative, schema-defined city model queried through a text-to-Cypher
mechanism, not a corpus of retrieved text fragments. Recent work questions the
need for pre-built graphs in RAG altogether~\cite{chen_graphrag_2026}, while
others target the spatial-reasoning limitations of
LLMs~\cite{schneider_distrag_2025}. For city models specifically,
CityGPT~\cite{feng_citygpt_2025} fine-tunes a city-scale world model into the LLM.

Some recent related systems, like \mytoolname, expose CityGML
to natural-language querying. KCitychatBot~\cite{liu_kcitychatbot_2025} couples a multi-agent LLM pipeline with
a CityGML KG in Neo4j, constructed with a hierarchical mapping in
the 3DCityKG tradition~\cite{nguyen_phdthesis_2024}, on large-scale CityGML data. 
The chatbot generates Cypher but does not support spatial analysis, which its authors
defer to future work using external tools such as PostGIS, and it relies on an
embedding model for intent extraction.
\citet{kanna_llminteraction_2025,kanna_llmenrichment_2025}, by contrast, support
spatial and temporal querying and 3D~Tiles visualization, but store the city
model in the relational 3DCityDB and translate user questions into SQL, supplying the
LLM with a multi-table relational schema serialized as several hundred lines of
context. \mytoolname\ occupies the gap between these: it queries a compact,
spatially indexed CityGML property graph through text-to-Cypher, enabling spatial
predicates such as proximity, intersection, and containment to be combined with semantic
traversal in a single query. The schema is sufficiently compact to fit within 
the model’s context and does not require embeddings on the query path. 

\section{Graph Schema and CityGML Mapping}\label{sec:graph_schema}

We present the graph schema employed by \mytoolname\ 
and formalize the transformation rules used to map a CityGML dataset into an LPG.

\subsection{Design Principles}
The schema is guided by five principles.

\subsubsection{Directed acyclic structure.} \mytoolname\ represents the city as a directed acyclic graph (DAG), with edge directions aligned to CityGML semantics. Aggregation and composition (``part-of'') relations consistently point from higher-level entities to their constituents (such as from building down to surface, polygon, and ring). This orientation allows queries, including those generated by LLMs, to infer traversal direction quickly based on local node and edge labels, independently from global schema. Although CityGML may permit directed cycles (e.g., upward XLinks), such cases are rare and absent in our datasets. Enforcing acyclicity thus preserves observed information while guaranteeing termination of recursive traversals.

\subsubsection{Compactness.} 
\begin{figure}[t]
  \centering
  \def\vstep{0.8} %
  \definecolor{scFeaB}{RGB}{0,105,92}    \definecolor{scFeaF}{RGB}{224,242,241} %
  \definecolor{scGeoB}{RGB}{46,125,50}   \definecolor{scGeoF}{RGB}{232,245,233} %
  \definecolor{scWrapB}{RGB}{122,122,122}\definecolor{scWrapF}{RGB}{246,246,246} %
  \includegraphics{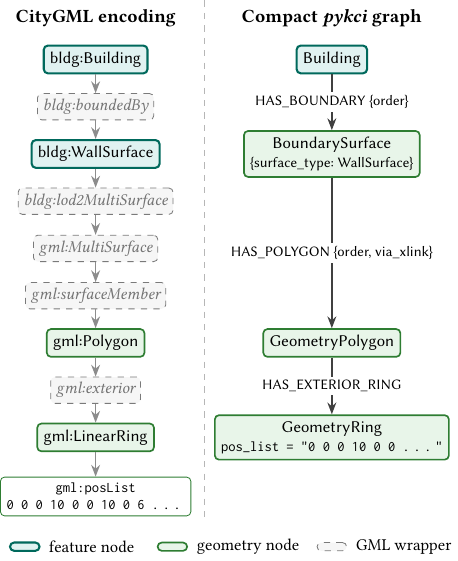}
  \caption{An example of CityGML encoding (left) and compact \mytoolname\ graph (right) of a building wall surface. The CityGML serialization nests the content through ten elements, while \mytoolname\ captures the same content in three hops. Each wrapper is absorbed into the spanning edge. Edge properties are shown in curly brackets. Verbatim coordinates are stored in \texttt{pos\_list}.}
  \Description{Diagram comparing CityGML encoding and compact graph representation of a building wall surface.}
  \label{fig:wrapper}
\end{figure}

Traversal paths are shortened by mapping only semantically meaningful CityGML elements to nodes, centralizing them around top-level feature nodes (like buildings), and collapsing wrappers in the CityGML serialization. This contrasts with hierarchy-faithful
mappings such as 3DCityKG~\cite{nguyen_phdthesis_2024} and significantly reduces both query hops and
the amount of context an LLM must process, as illustrated in~\Cref{fig:wrapper}. By a wrapper, we refer to an intermediate GML element (e.g., \texttt{gml:surfaceMember}, \texttt{gml:exterior}) that only groups or nests geometry and carries neither content nor an identifier of its own.

\subsubsection{Losslessness.} Every identifier, thematic attribute, ordering, XLink indicator, and coordinate string is preserved, enabling exact reconstruction of the original CityGML (\Cref{sec:export_from_graph}). While compactness and losslessness appear conflicting, since faithful reconstruction tends to require materializing every wrapper, we resolve this with a single rule: information needed only for serialization is stored as node or edge properties, not as additional nodes.

This yields three consequences. (1) Semantic content (e.g., city objects, rooms, openings, boundary surfaces, and geometry hierarchies) is represented as nodes and edges, with their order among sibling  elements stored as the edge property \texttt{order}.
(2) XLink references are modeled as explicit edges, annotated with a property \texttt{via\_xlink} set to \texttt{true}, to support CityGML reconstruction.
Thus, wrappers are reduced to edge properties, preserving one-hop references. 
(3) Irregular or open-ended leaf data, such as \texttt{pos\_list}, xAL addresses, metadata, and Application Domain Extension (ADE) content, is stored verbatim as string properties and reattached during export rather than modeled structurally.

\subsubsection{Dataset idempotency.} Ingestion uses Cypher’s \texttt{MERGE} clause, which combines \texttt{MATCH} and \texttt{CREATE}: it first searches for a node pattern and creates it only if absent. As a result, re-ingesting a dataset does not produce duplicates, and shared geometries are merged into single nodes. This is particularly beneficial for XLink resolution. During mapping, an XLink may refer to an element that has not yet been encountered, which would require collecting all references first before matching. With \texttt{MERGE}, referenced nodes are created on first mention and enriched later, eliminating the need for deferred-reference lists and post-processing.

\subsubsection{Spatial indexing.} Location queries, such as retrieving features within a bounding box or near a point, would otherwise require a full label scan with a per-node coordinate comparison. To avoid this, in addition to Neo4j's built-in point index on feature centroids, an R-tree spatial index is employed. It registers every top-level feature that has a \texttt{gml:boundedBy} envelope (\texttt{Building} and all \texttt{CityObject} subtypes) by its 2D bounding box. The box is stored in WKT format as a node property, which can be parsed by the R-tree layer. 

\begin{figure*}[t]
  \centering
  \definecolor{scCtxB}{RGB}{69,90,100}   \definecolor{scCtxF}{RGB}{236,239,241} %
  \definecolor{scFeaB}{RGB}{0,105,92}    \definecolor{scFeaF}{RGB}{224,242,241} %
  \definecolor{scTheB}{RGB}{239,108,0}   \definecolor{scTheF}{RGB}{255,243,224} %
  \definecolor{scGeoB}{RGB}{46,125,50}   \definecolor{scGeoF}{RGB}{232,245,233} %
  \includegraphics{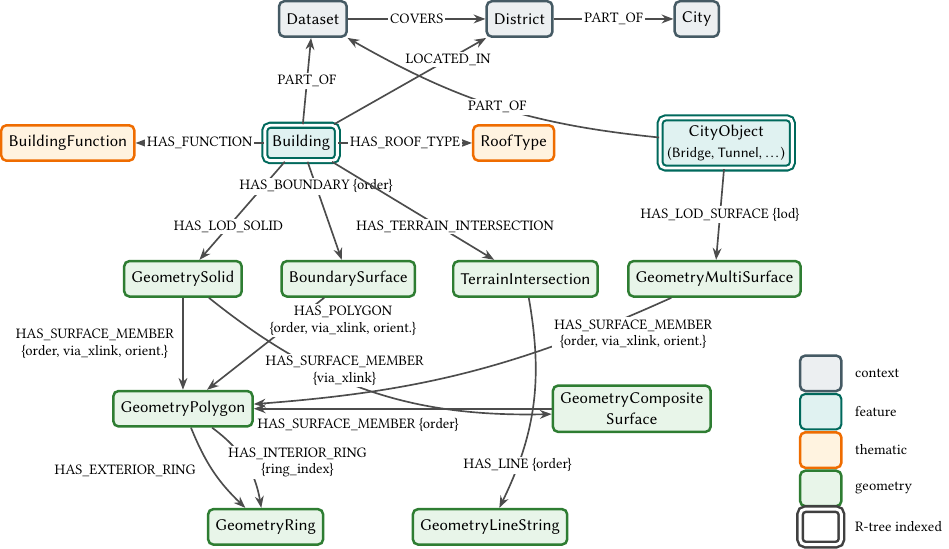}
  \caption{The \mytoolname\ core graph schema, comprising five node categories: context (gray), feature (dark green), thematic (orange), and geometry (green); interior or nested nodes are omitted for clarity due to space constraints.
  Feature nodes drawn with a double border (\texttt{Building}, \texttt{CityObject}) contain a 2D bounding box and are spatially indexed using an R-tree.
  Edge properties preserve information required for lossless reconstruction from the graph to CityGML.
  Depending on the use case and dataset, selected semantically significant attributes can be promoted to thematic nodes (such as building function and roof type).}
  \Description{A schematic graph showing five node categories: context, feature, thematic, and geometry, with color coding and omitted interior nodes for clarity.}
  \label{fig:schema}
\end{figure*}

\begin{figure*}
    \definecolor{gmNode}{RGB}{36,161,75}    %
    \definecolor{gmProp}{RGB}{33,118,222}   %
    \definecolor{gmVbat}{RGB}{240,150,20}   %
    \definecolor{gmEdge}{RGB}{156,44,185}   %
    \definecolor{gmNA}{RGB}{120,120,120}    %
    
    \providecommand{\sNode}{\tikz[baseline=-0.5ex]{\fill[gmNode] (0,0) circle (0.62ex);}}
    \providecommand{\sProp}{\tikz[baseline=-0.5ex]{\fill[gmProp] (-0.56ex,-0.56ex) rectangle (0.56ex,0.56ex);}}
    \providecommand{\sPropV}{\tikz[baseline=-0.5ex]{\fill[gmVbat] (0,0.66ex)--(0.66ex,0)--(0,-0.66ex)--(-0.66ex,0)--cycle;}}
    \providecommand{\sEdge}{\tikz[baseline=-0.5ex]{\draw[gmEdge,line width=1.3pt,line cap=round] (-0.6ex,0)--(0.6ex,0) (0,-0.6ex)--(0,0.6ex);}}
    \providecommand{\sNA}{\tikz[baseline=-0.5ex]{\draw[gmNA,line width=1.3pt,line cap=round] (-0.45ex,-0.45ex)--(0.45ex,0.45ex) (-0.45ex,0.45ex)--(0.45ex,-0.45ex);}}
    
    \providecommand{\gm}[1]{\llap{\makebox[1.7ex][c]{\smash{#1}}\kern0.45em}}

    \centering
    \includegraphics{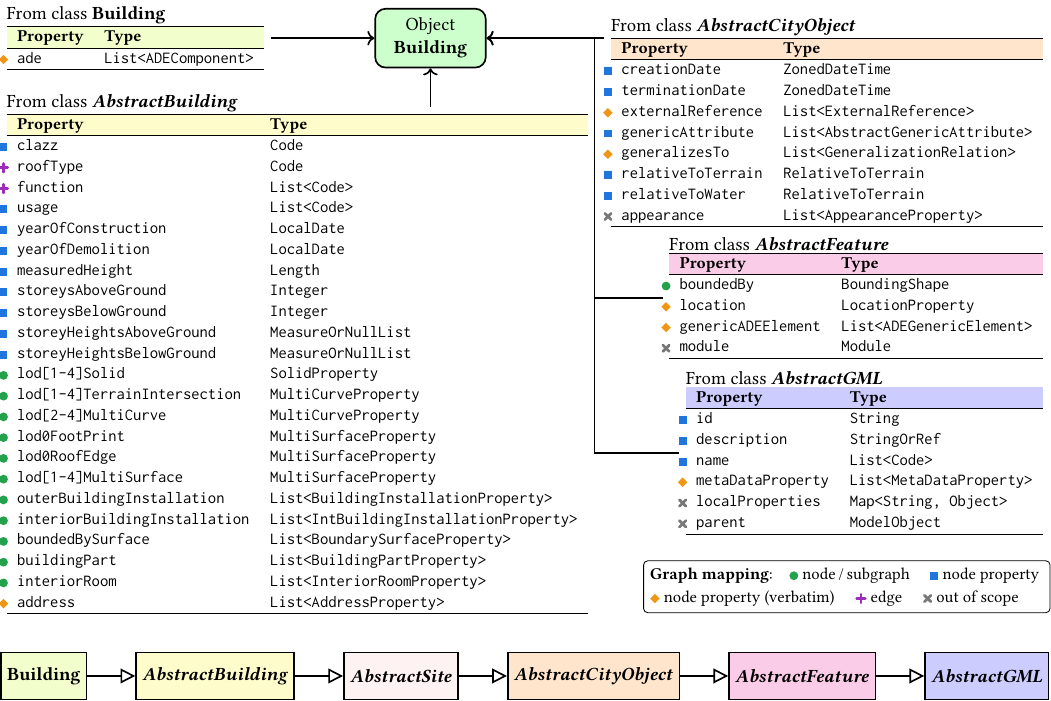}
    \caption{Content composition of a CityGML \textit{Building} object. Properties are grouped by their defining class, with the class hierarchy shown at the bottom. Property types follow the Java bindings of the \texttt{citygml4j} parser. All substantial content can be mapped to graph using \mytoolname's own graph schema: \protect\sNode~mapped to a dedicated node or subgraph, \protect\sProp~stored as a node property, \protect\sPropV~stored verbatim as an XML subtree (opaque to queries), and \protect\sEdge~represented as an edge to a shared node. \protect\sNA~denotes elements outside the scope of this work (e.g., parser artifacts or the excluded \textit{Appearance} module). Adapted from~\citet{nguyen_phdthesis_2024}.}
    \label{fig:citygml_model}
\end{figure*}

The core graph schema is summarized in \Cref{fig:schema}. \Cref{fig:citygml_model} illustrates how \texttt{Building} elements are mapped to graph entities using the schema and mapping rules defined in this work.

\subsection{Mapping CityGML to Graph}

The schema defines five node groups.
(1)~\emph{Feature nodes} represent CityGML city objects, including top-level features like \texttt{Building}, \texttt{Bridge}, and \texttt{Tunnel}. Notably, the model supports recursive structures through the \texttt{Building\-Part} node, which is connected to a \texttt{Building} via a \texttt{HAS\_BUILDING\_PART} edge and reuses the full \texttt{Building} geometry modeling schema.
(2)~\emph{Interior feature nodes} extend the schema to support LoD3--4 representations, including \texttt{Room}, \texttt{Building\-Installation}, \texttt{Building\-Furniture}, and \texttt{Opening} (covering doors and windows).
(3)~\emph{Geometry nodes} represent the hierarchical spatial decomposition of objects. These include \texttt{Geometry\-Solid}, \texttt{Geometry\-Multi\-Surface}, \texttt{Geometry\-Composite\-Surface}, \texttt{Boundary\-Surface}, \texttt{Geometry\-Polygon}, \texttt{Geometry\-Ring}, \texttt{Geometry\-LineString}, and \texttt{Terrain\-Intersection}.
(4)~\emph{Thematic nodes} capture semantically significant attributes, particularly code-list values such as \texttt{Building\-Function} and \texttt{Roof\-Type}. Due to their potentially high query frequency, these values are modeled as standalone nodes rather than node properties. Identical values are stored once and shared across all referencing features (via \texttt{MERGE}), allowing for explicit, queryable relationships. For instance, selecting a \texttt{Building\-Function} node indicating residential use enables efficient retrieval of all associated buildings via their connecting edges.
(5)~\emph{Context nodes} (e.g., \texttt{Dataset}, \texttt{District}, \texttt{City}) represent information within administrative and provenance contexts. The \texttt{City} node is derived from the \texttt{City\-Model} element in the CityGML document, while \texttt{Dataset} and \texttt{District} nodes are introduced based on available metadata of the dataset.

Relationships are divided into three categories.
(1)~\emph{Decomposition} edges encode geometric and part--whole containment, including 
\texttt{HAS\_\-LOD\_\-SOLID}, \texttt{HAS\_\-BOUNDARY}, \texttt{HAS\_\-POLYGON},
\texttt{HAS\_\-SURFACE\_\-MEMBER}, \texttt{HAS\_\-EXTERIOR\_\-RING},
\texttt{HAS\_\-INTERIOR\_\-RING}, \texttt{HAS\_\-LOD\_\-SURFACE}, and \texttt{HAS\_\-LINE}.
Interior nodes have their own edges, including \texttt{HAS\_\-INTERIOR\_\-ROOM}, \texttt{HAS\_\-OPENING},
and \texttt{HAS\_\-BUILDING\_\-PART}. 
Edges may contain properties reserved for CityGML reconstruction: \texttt{order} (position among siblings), \texttt{via\_\-xlink}
(XLink indicator), \texttt{orientation} (for \texttt{gml:Orientable\-Surface}), and \texttt{ring\_\-index} (for position of interior rings in a polygon).
(2)~\emph{Classification} edges like \texttt{HAS\_FUNCTION} and \texttt{HAS\_ROOF\_TYPE}
link a feature to its shared thematic nodes.
(3)~\emph{Context} edges (e.g., \texttt{PART\_OF}, \texttt{LOCATED\_IN}, and \texttt{COVERS})
attach features to their dataset, district, and city. 

\subsection{Export from Graph}\label{sec:export_from_graph}
\mytoolname\ serves as an operational data store and supports export to two formats. For interactive visualization, geometry is exported as an OGC 3D Tiles~1.1~\cite{cozzi_3dtiles_2023} dataset (\texttt{tileset.json} with binary glTF tiles) and rendered in CesiumJS. The coordinate reference system (CRS) and geoid undulation are derived from the node properties \texttt{srsName} in the graph, surfaces are triangulated,
and per-feature identifiers are embedded in the GLB visualization files. This tileset provides the spatial context in \Cref{fig:teaser}.

For data exchange, the export reconstructs a schema-valid CityGML~2.0 document covering all ingested modules and levels of detail. Since identifiers, ordering, typed attributes, code spaces, coordinate strings, and XLink relationships are preserved during ingestion, the output restores geometry, semantics, and metadata, including LoD geometries, boundary surfaces, openings, building parts, installations, and interiors. 
The result is semantically complete with respect to the source (it contains every element, attribute, text value, and coordinate of the input), enabling round-trip workflows. 
The appearance module (\texttt{app:Appearance}) is excluded.

\section{Evaluation Results}\label{sec:evaluation}

We define a round trip as \emph{lossless} in terms of semantic completeness: every element, attribute, leaf text value, and coordinate in the source must appear in a valid CityGML export, although encoding (e.g., \texttt{gml:pos} vs.\ \texttt{posList}), element order, and whitespace may differ. 
To measure this property, an automated
census parses the source and the export and reduces each document to four
multisets: element local-names, \texttt{(attribute, value)} pairs,
\texttt{(element, text)} pairs, and coordinate tokens. A coordinate token is
an individual scalar value parsed from \texttt{gml:pos} or \texttt{gml:posList}.
The export covers the source if and only if every source multiset is contained in
the corresponding export multiset. 
We apply the census on
two axes: all LoDs (\Cref{subsec:census_lod}) and the thematic feature
modules (\Cref{subsec:census_modules}).

\subsection{Mapping All Levels of Detail (LoD0--4)}\label{subsec:census_lod}

\begin{figure}[h]
  \centering
  \includegraphics[width=0.9\linewidth]{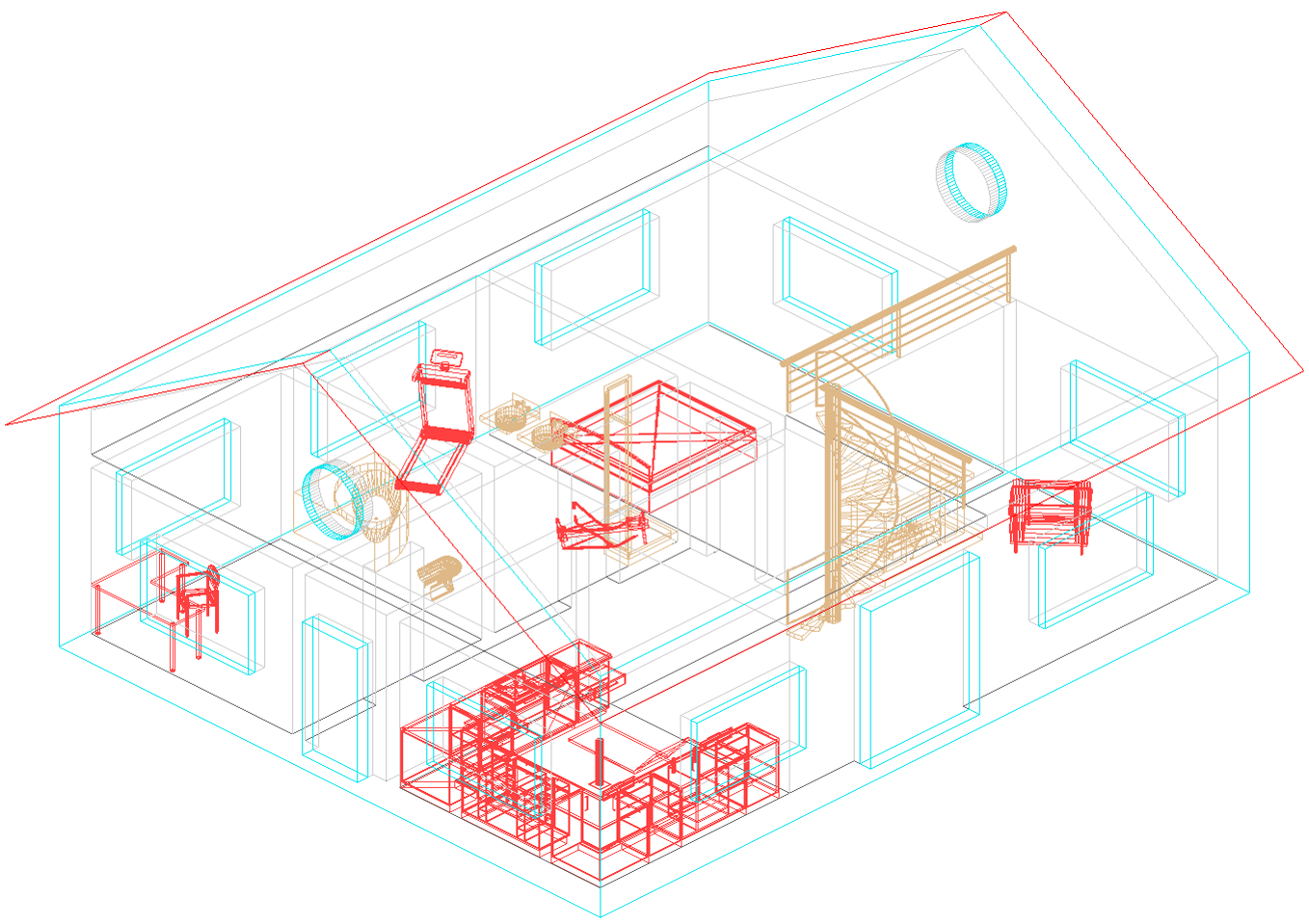}
  \caption{Visualization of the FZK-Haus CityGML dataset in LoD4. Rendered in KITModelViewer.}
  \label{fig:fzkhaus_lod4}
\end{figure}

We use the FZK-Haus reference building published by KIT~\cite{kit_fzkhaus_2021} in all five LoD0--4. 
These datasets contain per-vertex coordinates, 
identifiers, composite and orientable surfaces, and, at LoD4, a complete interior
of rooms, furniture, and installations (see~\Cref{fig:fzkhaus_lod4}). 
For each LoD, we execute the pipeline end to
end, ingesting the source into Neo4j and exporting it back to CityGML, and then
census the source against the export. \Cref{tab:census} counts every source
instance by category and reports the fraction that the export covers. The source
grows by more than three orders of magnitude across the LoDs, from \num{117}
instances at LoD0 to \num{275366} at LoD4, where the interior elements dominate. The
mapping covers \qty{100}{\percent} of the instances at every LoD. 

\begin{table}[t]
  \caption{Round-trip census on the FZK-Haus datasets in LoD0--4. Source
instances counted in four multisets (element names, attributes, text, coordinate
tokens). The export covers every instance at all LoDs (100 \%).
}
  \label{tab:census}
  \begin{tabular}{@{}lrrrrrr@{}}
    \toprule
    LoD & Elem. & Attr. & Text & Coord. & Source total & Coverage \\
    \midrule
    0 & \num{49}    & \num{16}  & \num{22}  & \num{30}     & \num{117}     & \qty{100}{\percent} \\
    1 & \num{65}    & \num{16}  & \num{22}  & \num{90}     & \num{193}     & \qty{100}{\percent} \\
    2 & \num{112}   & \num{23}  & \num{30}  & \num{111}    & \num{276}     & \qty{100}{\percent} \\
    3 & \num{1036}  & \num{36}  & \num{43}  & \num{4191}   & \num{5306}    & \qty{100}{\percent} \\
    4 & \num{59699} & \num{208} & \num{194} & \num{215265} & \num{275366}  & \qty{100}{\percent} \\
    \bottomrule
  \end{tabular}
\end{table}

\subsection{Mapping the Thematic Feature Modules}\label{subsec:census_modules}

\begin{figure}[h]
  \centering
  \includegraphics[width=\linewidth]{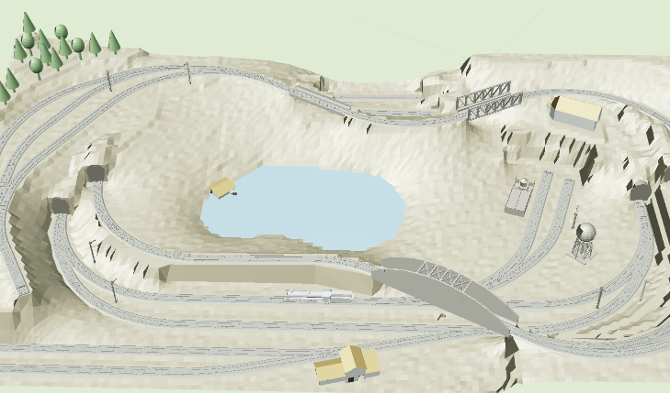}
  \caption{3D Tiles visualization of the Railway Scene dataset in LoD3 with ten CityGML modules.}
  \label{fig:railway_scene_lod3}
  \Description{Visualization of the Railway Scene dataset in LoD3 with ten CityGML modules.}
\end{figure}

To evaluate mapping of thematic modules beyond buildings, we employ the Railway Scene dataset in LoD3~\cite{haefele_railway_2015}, which packs
\num{52} top-level features from ten CityGML modules into a single scene (see~\Cref{fig:railway_scene_lod3}). These include
the modules \textit{Bridge} and \textit{Tunnel} with nested construction elements and installations, a native Triangulated Irregular Network (TIN) relief, XLink-based group membership, and generic city objects. We ingest the
whole scene, export it, and census the source against the export over the entire
document. \Cref{tab:census_railway} lists the modules and the distinctive
elements that each contributes. A check mark indicates that the module's content
has been mapped within the fully covered scene. In total, the census counts
\num{1182102} source instances (\num{243254} elements, \num{487} attributes,
\num{1194} text values, and \num{937167} coordinate tokens), and the export covers
all of them (\qty{100}{\percent}). Ten CityGML~2.0 thematic feature
modules appear in this scene. The eleventh module, \textit{LandUse}, is supported by the same surface-only geometry mapping as \textit{Transportation} but is not available in any of the evaluation datasets.
The only non-feature module \textit{Appearance} is out of
scope by design. 

\begin{table}[t]
  \caption{Round-trip census on the Railway Scene LoD3: 52 features across
ten CityGML modules. A check mark (\checkmark) indicates a covered module. 
The lower block provides whole-scene instance counts, all covered. 
The module \textit{Appearance} is excluded and \textit{LandUse} is not available in the dataset.}
  \label{tab:census_railway}
  \begin{tabularx}{\columnwidth}{@{}lrXc@{}}
    \toprule
    Module & Feat. & Distinctive elements & Cov. \\
    \midrule
    \textit{Building}          & 3  & openings, installations & \checkmark \\ %
    \textit{Bridge}            & 4  & nested con. ele., installations     & \checkmark \\
    \textit{Tunnel}            & 4  & nested installations, closure          & \checkmark \\
    \textit{Transportation}    & 10 & surface-only LoD railway             & \checkmark \\
    \textit{Vegetation}        & 15 & solitary obj., \texttt{lodN} geometry        & \checkmark \\
    \textit{CityFurniture}     & 11 & \texttt{lodN} geometry                          & \checkmark \\
    \textit{WaterBody}         & 1  & water boundary surfaces                         & \checkmark \\
    \textit{Relief}            & 1  & native triangulated surfaces      & \checkmark \\ %
    \textit{CityObjectGroup}   & 1  & XLink group membership                          & \checkmark \\
    \textit{GenericCityObject} & 2  & typed generic attributes                        & \checkmark \\
    \midrule
    \multicolumn{4}{@{}l}{\emph{Whole-scene instances covered:}} \\
    Elements   & \multicolumn{2}{r}{\num{243254}}  & \qty{100}{\percent} \\
    Attributes & \multicolumn{2}{r}{\num{487}}     & \qty{100}{\percent} \\
    Text       & \multicolumn{2}{r}{\num{1194}}    & \qty{100}{\percent} \\
    Coords     & \multicolumn{2}{r}{\num{937167}}  & \qty{100}{\percent} \\
    Total      & \multicolumn{2}{r}{\num{1182102}} & \qty{100}{\percent} \\
    \bottomrule
  \end{tabularx}
\end{table}

\section{Benchmarking Results}\label{sec:benchmarking}

We evaluate the performance of \mytoolname\ against two open-source CityGML database mappers: 3DCityKG~1.0~\cite{nguyen_phdthesis_2024} (graph-based, Neo4j) and 3DCityDB~5.0~\cite{yao_3dcitydb5_2025} (relational, PostgreSQL/PostGIS). All three systems ingest the same $6.93\,\mathrm{GB}$ CityGML~2.0 LoD2 dataset of Hamburg~\cite{lgv_hhcitygml_2025}, comprising $388{,}267$ buildings recorded in 2025. 
To ensure a fair comparison, each system is executed in Docker under identical hardware conditions (8 logical CPU cores at 4.7--5.2~GHz and $20\,\mathrm{GB}$ RAM), reads the dataset via the same bind mount, and is benchmarked over three runs (run-to-run variance under $3\%$). We report mean mapping wall-clock time 
as well as the resulting database size and graph entity counts.

\begin{figure}[h]
  \centering
  \includegraphics{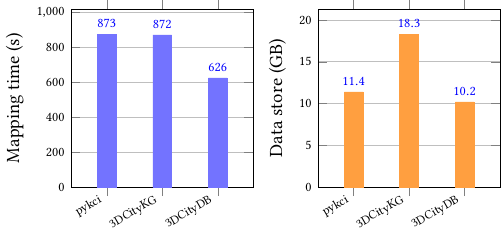}
  \caption{Mean mapping time (left) and database size (right) for the Hamburg CityGML 
  LoD2 dataset.}
  \label{fig:bench-perf}
\end{figure}

\begin{figure}[t]
  \centering
  \includegraphics{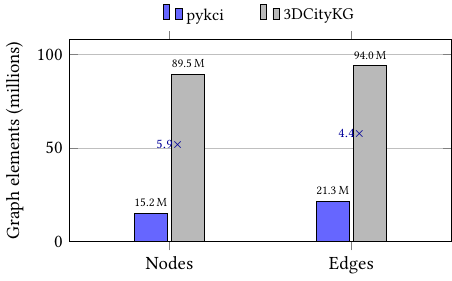}
  \caption{Graph compactness of \mytoolname\ compared to 3DCityKG for the same Hamburg CityGML LoD2 dataset.}
  \label{fig:stat_compactness}
\end{figure}

As shown in \Cref{fig:bench-perf}, \mytoolname\ maps the dataset in $873\,\mathrm{s}$ ($15\,\mathrm{min}$), statistically indistinguishable from 3DCityKG ($872\,\mathrm{s}$). The relational 3DCityDB is the fastest at $626\,\mathrm{s}$. 
\mytoolname\ is written in Python, while 3DCityKG and 3DCityDB are Java-based.
In terms of storage, the \mytoolname{} data store occupies $11.4\,\mathrm{GB}$, approximately \qty{40}{\percent} smaller than 3DCityKG at $18.3\,\mathrm{GB}$, while the relational 3DCityDB is slightly smaller at $10.2\,\mathrm{GB}$. Both \mytoolname{} and 3DCityDB include spatial indexing (R-tree and Generalized Search Tree (GiST), respectively), whereas 3DCityKG does not store a persistent spatial index. As shown in \Cref{fig:stat_compactness}, \mytoolname{} employs a condensed semantic schema, yielding nearly 6 times fewer nodes and more than 4 times fewer edges compared to 3DCityKG. For the same dataset, the relational 3DCityDB requires 27 million rows across 4 populated tables.

We evaluate the performance of \mytoolname, 3DCityKG, and 3DCityDB using 20 predefined natural-language questions, comprising 10 citizen-oriented queries (e.g., roof materials, tallest buildings) and 10 expert-oriented queries (e.g., spatial aggregates, solar scoring). The evaluation is conducted on the same Hamburg CityGML LoD2 model enriched with building roof materials (see \Cref{sec:usecases_enrichment}). We employ LLMs as text-to-query interfaces, including both local open-weight and commercial models. We evaluate three local models: Gemma 4, Mistral Medium 3.5, and Qwen3.5 (\Cref{tab:open-llm-detail}). Results obtained with Claude (Opus 4.8) are summarized in \Cref{tab:query-bench}.
The results show that \mytoolname{} produces the shortest queries on average (approximately 290 characters) and achieves the fastest net database query time at \SI{0.37}{\second} (wall-clock time excluding client startup overhead). Both the full list of natural-language questions and a detailed per-query performance evaluation across all databases can be found in \Cref{tab:query-list} and \Cref{tab:query-detail} in the Appendix, respectively.

\begin{table}[h]
  \centering 
  \setlength{\tabcolsep}{4pt}
  \caption{Query performance comparison of \mytoolname, 3DCityKG, and 3DCityDB across 20 predefined queries (10 citizen, 10 expert) on the Hamburg dataset enriched with roof materials. \mytoolname{} yields the most compact queries and the fastest net execution time. The full question set and detailed per-query analysis are provided in \Cref{tab:query-list} and \Cref{tab:query-detail} in the Appendix.}
  \label{tab:query-bench}

  \begin{tabular*}{\columnwidth}{@{\extracolsep{\fill}}lccc@{}}
    \toprule
    & \textbf{pykci} & \textbf{3DCityKG} & \textbf{3DCityDB} \\
    & (Neo4j) & (Neo4j) & (PostGIS) \\
    \midrule
    Query language                    & Cypher           & Cypher          & SQL \\
    Questions solved correctly$^{a}$   & 20/20            & 20/20           & 20/20 \\
    Schema-discovery queries$^{b}$     & \textbf{2}       & 7               & 3 \\
    Authoring difficulty (1--5)$^{c}$  & 2.1              & 3.9             & \textbf{2.0} \\
    Mean query length (chars)          & \textbf{$\sim$290} & $\sim$460     & $\sim$970\,$^{d}$ \\
    Net DB query time (mean)$^{e}$     & \textbf{0.37\,s} & 2.10\,s         & 2.62\,s \\
    Spatial index at query time       & R-tree           & none      & GiST \\
    \bottomrule
  \end{tabular*}

  \vspace{2pt}
  {\scriptsize\raggedright
   Results obtained using Claude (Opus 4.8) for text-to-Cypher translation and performance evaluation. 
   $^{a}$ Of 18 answerable questions; 2 of 20 are infeasible due to unavailable information (at most one roof material per building; no roof coverage field).\;
   $^{b}$ Exploratory queries needed to locate attributes before executing.\;
   $^{c}$ Mean over the 18 answerable queries (rubric, \Cref{tab:query-detail}).\;
   $^{d}$ Including a shared $\sim$745-char property-pivot Common Table Expression (CTE).\;
   $^{e}$ Wall-clock minus fixed client startup (cypher-shell JVM $\approx$1.45\,s,
   psql $\approx$0.2\,s).\;
   }
\end{table}

Our design limits hallucination as LLMs do not produce answers directly. Instead, they generate Cypher queries that are executable in Neo4j. Returned values are therefore grounded in the graph store, while the generated queries are auditable and reproducible. 
Of the 18 answerable queries (\Cref{tab:query-list}), six require multiple analytical steps (rubric in \Cref{tab:query-detail}), such as combining semantic and geometric information across buildings. Five 
are explicitly spatial, involving bounding boxes, grid-based aggregation, and estimation of floor-area ratio. Claude responded to all 18 questions correctly, with 16 solved on the first attempt. It correctly identified the two deliberately unanswerable questions (C9 and E7), which request information unavailable in the KG. 
Spatial tasks are the main challenge for local open-weight models despite the available spatial indexes and stored bounding boxes (C10 and E2). 
This evaluation suite is intended as a demonstration of coupling LLMs with our compact KGs. In related work~\cite{nguyen_authenticity_2026}, we provide a more comprehensive, broader query benchmark for large-scale knowledge stores generated and enriched by~\mytoolname.

\section{Use Cases: Graph-based Enrichment}
\label{sec:usecases_enrichment}

This section demonstrates how \mytoolname\ can be used as a knowledge base to support real-world workflows, specifically for enriching city models with additional data. This includes roof material classification and the extraction of facade openings from imagery. To illustrate this, we use the same Hamburg KG introduced in~\Cref{sec:benchmarking}.

\subsection{Roof Material Enrichment}
\begin{figure}[h]
  \centering
  \includegraphics[width=\linewidth]{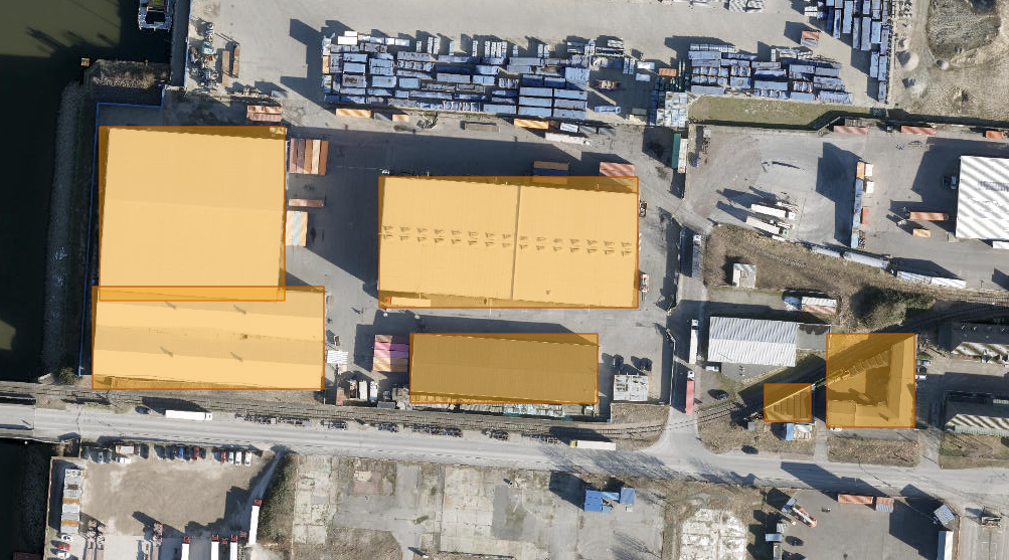}
  \caption{Visualization of enriched metal roof material attributes overlaid on the Esri World Imagery basemap and queried from \mytoolname\ using an LLM.}
  \label{fig:enriched_roofmats}
\end{figure}

We attach deep learning based roof material predictions from a companion GeoJSON dataset, following the methodology of \citet{lukas_roofmaterials}, to almost half of the 388{,}267 \texttt{Building} nodes.
The \mytoolname\ KG remains compact, with only five additional nodes representing the roof material classes. Buildings with assigned roof material are linked to these nodes via \texttt{HAS\_ROOF\_MATERIAL} relationships.
Based on the enriched roof materials, further analysis and visualization can be performed. For example, these properties can be mapped alongside orthophotos to visually verify the prediction of such attributed materials, as illustrated in Figure~\ref{fig:enriched_roofmats}.

\subsection{Facade Opening Enrichment (LoD3)}
\begin{figure}[h]
  \centering
  \includegraphics[width=\linewidth]{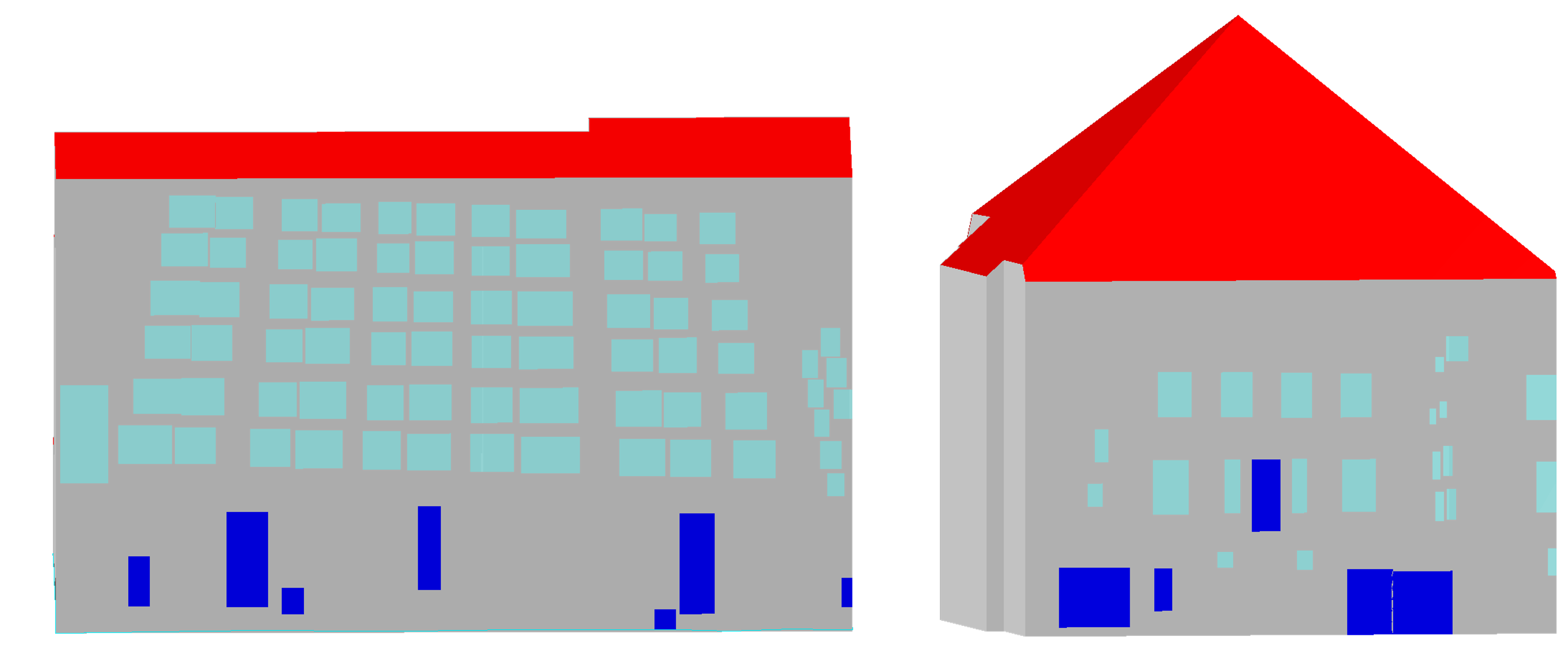}
  \caption{Semantic enrichment for buildings with diverse morphologies: roof top materials (red, right), tar paper roofs (red, left), with windows (cyan) and doors (blue). The resulting roof and facade semantics are stored within~\mytoolname.}
  \Description{Visualization showing buildings with different roof materials and colored highlights for windows and doors.}
  \label{fig:material_openings_pixi_integration}
\end{figure}

For 17 buildings in Hamburg, we derive their facade openings (i.e., windows and doors), including position and shape, from RGB segmentation masks. \Cref{fig:material_openings_pixi_integration} visualizes such detected windows in cyan and doors in blue. Our approach maps bounding boxes to normalized UV coordinates in $[0,1]^2$, following the CityGML convention, where $v = 0$ corresponds to ground level. Facade openings are represented as nodes, each defined by normalized $u$ and $v$ coordinates and a semantic type. Similar to roof materials, these opening nodes are connected to buildings via \texttt{HAS\_FACADE\_OPENING} edges.

\subsection{Downstream Applications in Urban Context}

\emph{Greening planning and heat island mitigation.} We aggregate dark roofing materials, mainly tar paper and metal, over a \SI{2}{\kilo\metre} grid to derive a dark-roof fraction per cell. The highest value, \SI{49}{\percent}, occurs in HafenCity with high concentration of industrial buildings. 
\citet{lukas_roofmaterials} and~\citet{kanna2026semantic} enrich CityGML datasets with footprint-guided roof materials and derive scene-specific land surface temperature (LST) estimates. Their framework supports cool-roof and green-roof retrofit evaluation, reporting simulated citywide cooling of about \SI{0.75}{\kelvin} in Hamburg, \SI{0.62}{\kelvin} in Madrid, and \SI{0.22}{\kelvin} in Paris. As all enriched roof materials and facade openings are linked directly to buildings through the \mytoolname\ graph schema, climate-relevant data can be retrieved and processed through a single one-hop graph traversal.

\emph{Solar scoring.} Another application of the enriched semantic property graph is photovoltaic (PV) suitability assessment. We compute a composite score based on roof type, material, slope, and building height to rank buildings for solar installations. Flat glass or concrete roofs at least \SI{5}{\metre} high receive the highest scores. Because buildings with the same roof material are connected, queries over material classes can be executed efficiently. This allows direct use of material properties in graph-based ranking queries, for example through a single lookup within a \texttt{reduce()} expression. 

\emph{Facade typology.}
Texture coordinates UV support facade-level analysis without complete 3D geometry. The window-to-door ratio can indicate building function, with higher values typically linked to residential buildings. Dividing the vertical coordinate $v$ into intervals approximates opening density across storeys. These metrics support daylight availability assessment, occupancy estimation, and energy-demand modeling. Resulting daylight and noise simulations help identify suitable locations for urban densification by balancing development potential and environmental quality. We enrich the data required for these analyses in a single~\mytoolname\ KG. \Cref{tab:enrichment} summarizes this section. 

\begin{table}[h]
  \caption{Reconstructed semantic knowledge integrated to the \mytoolname\ graph and the analyses it enables.}
  \label{tab:enrichment}
  \begin{tabular}{@{}lll@{}}
    \toprule
                & Roof material            & Facade openings        \\
    \midrule
    Source      & predicted GeoJSON        & reconstructed CityGML     \\
    Pattern     & \texttt{HAS\_ROOF\_MATERIAL} & \texttt{HAS\_FACADE\_OPENING} \\
    Nodes/edges & 5 / 195{,}395            & 499 / 499              \\
    Runtime     & 15\,s                    & 7.8\,s                 \\
    Applications    & UHI, solar, V2X        & typology, daylight \\
    \bottomrule
  \end{tabular}
\end{table}

\section{Conclusion}\label{sec:conclusion}

We presented \mytoolname, an open-source, end-to-end Python system that maps CityGML~2.0 datasets into a compact Neo4j-based urban KG, supports export to OGC 3D Tiles and lossless reconstruction to CityGML, and enables natural-language querying via a model-agnostic text-to-Cypher interface. The graph schema covers all thematic feature modules and LoDs. It integrates semantic as well as spatial point and R-tree indexing.  
We evaluated and demonstrated the system on Hamburg CityGML LoD2 data, including complex tasks such as identifying roof surfaces suitable for greening, with results grounded in exact graph data.

\mytoolname\ offers several advantages. Its compact schema reduces LLM context size and enables exact query answering via Cypher rather than approximate similarity search, allowing small open-weight models to handle routine queries. The backend is interchangeable, enabling deployment across cities without retraining. Furthermore, the knowledge base remains local and private, requiring no tokenization or embedding of city data during query processing. 
However, the approach relies on an online database as external memory and abstracts away CityGML's data model structure. While this improves semantic query efficiency, it reduces suitability for fine-grained structural comparison of city objects, such as sub-graph matching, which remains better supported by hierarchy-preserving mappings such as 3DCityKG. 

Within UDTs, \mytoolname\ serves as a knowledge base accessible to both humans and machines. It provides a centralized platform for storing and integrating heterogeneous urban data. 
In a follow-up study~\cite{nguyen_authenticity_2026}, we extend \mytoolname\ to construct and enrich KGs of five different cities on three continents. These graphs combine authoritative CityGML LoD1-3 with crowd-sourced OSM data. In the future, we plan to also support CityGML 3.0 as more datasets become available. 
The project is published at \url{https://github.com/hcu-cml/pykci}. 

\begin{acks}
This research was partially supported by the project ``Next Generation City Networking'' (Grant No. 19DZ24004) at the Hanseatic Wireless Innovation Competence Center (HAWICC). The project is funded by the Federal Ministry of Transport via the German Center for Future Mobility (DZM).
\end{acks}

\bibliographystyle{ACM-Reference-Format}
\bibliography{sample-base}

\appendix

\section{Query Evaluation using LLMs}\label{app:query_evaluation}

This appendix includes: (1) the full list of verbatim natural-language queries used throughout this paper (\Cref{tab:query-list}); (2) a per-query performance analysis of Claude (Opus 4.8) on \mytoolname, 3DCityKG, and 3DCityDB; and (3) a per-query performance overview of local open-weight LLMs (Gemma 4, Mistral Medium 3.5, and Qwen3.5) on \mytoolname.

\begin{table*}[h]
  \centering 
  \small
  \caption{Verbatim natural-language queries used for benchmarking. The first ten queries (C1–C10) reflect a city-citizen perspective, while the remaining ten (E1–E10) represent data and GIS expert perspectives. The queries cover various difficulty levels. Queries C9 and E7 exceed the available data in the underlying knowledge base and are included to evaluate whether the LLMs can correctly detect and handle such limitations.}
  \label{tab:query-list}
  \begin{tabular}{@{}l p{12.4cm} l@{}}
    \toprule
    \# & Natural-language question (verbatim) & Capability\\
    \midrule
    \multicolumn{3}{@{}l}{\emph{Citizen perspective}}\\
    C1 & Hamburg at a glance: How many buildings are in the dataset, how tall are they on average, and how many storeys does a typical Hamburg building have? & Global aggregate\\
    C2 & What is Hamburg's most common roof material? Shows the distribution of predicted roof materials across all buildings, with percentage share. & Attribute aggregate\\
    C3 & What are the 20 tallest buildings in Hamburg? Good to spot landmarks, towers, and high-rises. & Top-$k$ ranking\\
    C4 & Which roof shapes are most popular in Hamburg? Flat, gable, hip, mono-pitch, etc. Show the architectural character of different neighborhoods. & Code$\to$label grouping\\
    C5 & Solar panel potential: Flat-roofed buildings taller than 5 m. Flat roofs on larger buildings are the easiest surfaces for solar installations. This query finds them and ranks by roof area proxy (height). & Multi-attr.\ filter + rank\\
    C6 & How many buildings are there per storey count? Show whether Hamburg is a city of single-family houses or dense apartment blocks. & Histogram\\
    C7 & Which buildings have glass roofs? Glass roofs typically indicate modern commercial buildings, train stations, shopping centers, or public spaces. & Attribute filter\\
    C8 & Hamburg high-rises: buildings taller than 50 meters. How many skyscrapers and tower blocks does Hamburg have? & Scalar filter\\
    C9 & Buildings with mixed roof materials: Some buildings have more than one material on the roof, which is interesting for understanding complex roof landscapes or renovation history. & \textit{Infeasible} (1 material/bldg)\\
    C10 & Buildings in HafenCity, Hamburg's modern harbor district: Filter by bounding box in ETRS89 UTM32 coordinates (meters). Shows roof materials and height, great for exploring a specific neighborhood's architectural character. & Spatial bbox filter\\
    \addlinespace
    \multicolumn{3}{@{}l}{\emph{Expert perspective}}\\
    E1 & Building typology: Height and storey profile per use type. Urban planners use this to understand the vertical density profile of each function class: residential, commercial, industrial, etc. Building type is extracted from the auto-generated description field. & Semantic grouping\\
    E2 & 1 km spatial grid: Building density, height, and dominant roof material. Breaks Hamburg into 1 km $\times$ 1 km cells and summarize each. Useful for density zoning analysis and morphological mapping. Only cells with $\geq$ 30 buildings are shown to filter sparse rural edges. & Spatial grid + mode\\
    E3 & Solar potential scoring: Ranked list of best candidates. Composite score based on: roof shape (flat = highest), material suitability, height above 10 m (clears most trees/obstacles), and multi-storey (higher energy demand justifies installation). Score range: 0 (unsuitable) to 7 (excellent). & Composite scoring\\
    E4 & Floor-to-floor height anomaly detection: Data quality flag. A typical storey is 3--4 m. Values outside 2--8 m indicate either special structures (industrial halls, towers, carports) or measurement errors in the source data, which is valuable for data quality auditing. & Derived ratio + anomaly\\
    E5 & Roof material distribution by building function. Cross-tabulation: for each use type, what fraction of buildings has each predicted roof material? Reveals whether material predictions correlate with known building typologies. Only types with $\geq$ 500 buildings included. & Cross-tabulation\\
    E6 & Urban heat island proxy: Dark roof fraction per 2 km grid cell. Tar paper and metal absorb significantly more solar radiation than concrete, tiles, or glass. High dark-roof density correlates with elevated surface temperatures (urban heat island effect). Only cells with $\geq$ 50 buildings with material predictions included. & Spatial grid + cond.\ \%\\
    E7 & Roof material prediction coverage completeness. The \texttt{material\_cov} values from the ML prediction indicate what fraction of the roof each material covers. A building with total coverage $<$ 0.9 has an unexplained portion, which is useful to assess prediction confidence and identify buildings that need re-inspection. & \textit{Infeasible} (no field)\\
    E8 & High-rise cluster detection: 500 m grid cells with multiple tall buildings. Tall buildings ($>$ 30 m) rarely stand alone: this identifies dense high-rise clusters useful for wind, shadow, and microclimate analysis. Each cell reports count, height stats, and sample building IDs. & Spatial grid + stats\\
    E9 & Floor area ratio (FAR) proxy per 1 km grid cell. FAR = total floor area / plot area. Here approximated as: gross floor area $\approx$ bbox\_footprint $\times$ storeys\_above\_ground. Aggregated to 1 km cells as a density planning indicator. High FAR cells indicate compact, high-intensity development. & Spatial grid + geometry\\
    E10 & Ground elevation profile: Flood risk and topographic context. Hamburg has areas below sea level (dikes, port basins) down to $-5$ m and elevated areas up to 88 m (Alstertal/Bergedorf ridges). This query combines ground\_z with building height to classify buildings by elevation zone, which is relevant for flood risk planning. & Elevation classification\\
    \bottomrule
  \end{tabular}
\end{table*}

\begin{table*}[h]
  \centering 
  \setlength{\tabcolsep}{3.5pt}
  \caption{Per-query performance of Claude (Opus 4.8) on \mytoolname, 3DCityKG, and 3DCityDB.}
  \label{tab:query-detail}
  \begin{tabular}{@{}ll cccc cccc cccc@{}}
    \toprule
    & & \multicolumn{4}{c}{\textbf{pykci} (Neo4j)} & \multicolumn{4}{c}{\textbf{3DCityKG} (Neo4j)} & \multicolumn{4}{c}{\textbf{3DCityDB} (PostGIS)} \\
    \cmidrule(lr){3-6}\cmidrule(lr){7-10}\cmidrule(lr){11-14}
    \# & Question & D & T & L & $t$ & D & T & L & $t$ & D & T & L$^{\ddagger}$ & $t$ \\
    \midrule
    C1  & City at a glance        & 1 & 1 & 151 & 0.30 & 3 & 1 & 242 & 3.84 & 1 & 1 &  871 & 3.24 \\
    C2  & Top roof material \%    & 2 & 2 & 255 & 0.25 & 4 & 1 & 368 & 2.52 & 2 & 1 &  901 & 3.66 \\
    C3  & 20 tallest buildings    & 1 & 1 & 206 & 0.16 & 3 & 1 & 347 & 0.89 & 1 & 1 &  881 & 2.49 \\
    C4  & Roof-shape histogram    & 2 & 1 & 449 & 0.31 & 3 & 1 & 460 & 0.76 & 2 & 1 & 1164 & 3.18 \\
    C5  & Flat roofs $>$5\,m      & 1 & 1 & 108 & 0.33 & 3 & 1 & 198 & 0.87 & 1 & 1 &  822 & 2.18 \\
    C6  & Storey histogram        & 1 & 1 & 134 & 0.25 & 2 & 1 & 141 & 0.22 & 1 & 1 &  845 & 3.00 \\
    C7  & Glass roofs             & 1 & 1 &  89 & 0.33 & 4 & 1 & 251 & 2.36 & 1 & 1 &  809 & 3.01 \\
    C8  & High-rises $>$50\,m     & 1 & 1 &  77 & 0.22 & 3 & 1 & 138 & 0.72 & 1 & 1 &  799 & 2.14 \\
    C9  & Mixed materials$^{\dagger}$ & -- & 1 & 198 & 0.23 & -- & 1 & 294 & 2.25 & -- & 1 &  908 & 0.18 \\
    C10 & HafenCity bbox          & 2 & 1 & 236 & 0.21 & 5 & 1 & 485 & 1.71 & 2 & 1 &  928 & 0.41 \\
    E1  & Typology by use type    & 2 & 1 & 252 & 0.98 & 4 & 1 & 364 & 1.53 & 2 & 1 &  943 & 2.43 \\
    E2  & 1\,km grid + dom.\ mat. & 4 & 2 & 611 & 0.43 & 5 & 1 & 1063 & 4.26 & 3 & 1 & 1131 & 2.78 \\
    E3  & Solar score 0--7        & 3 & 1 & 505 & 0.65 & 5 & 1 & 768 & 3.88 & 3 & 1 & 1135 & 2.60 \\
    E4  & Storey-height anomaly   & 2 & 1 & 191 & 0.31 & 3 & 1 & 252 & 0.94 & 2 & 1 &  866 & 2.24 \\
    E5  & Material $\times$ use   & 3 & 1 & 450 & 0.72 & 5 & 1 & 570 & 3.14 & 3 & 1 & 1147 & 3.46 \\
    E6  & Dark-roof / 2\,km cell  & 3 & 1 & 444 & 0.37 & 5 & 1 & 835 & 4.44 & 3 & 1 & 1103 & 4.03 \\
    E7  & Coverage compl.$^{\dagger}$ & -- & 1 & 190 & 0.10 & -- & 1 & 244 & 2.38 & -- & 1 &  862 & 2.75 \\
    E8  & High-rise clusters      & 3 & 1 & 414 & 0.21 & 5 & 1 & 738 & 1.81 & 3 & 1 & 1122 & 2.39 \\
    E9  & FAR / 1\,km cell        & 3 & 1 & 459 & 0.76 & 5 & 1 & 788 & 1.89 & 3 & 1 & 1087 & 3.93 \\
    E10 & Elevation zones         & 2 & 1 & 434 & 0.31 & 4 & 1 & 603 & 1.63 & 2 & 1 & 1136 & 2.23 \\
    \midrule
    \multicolumn{2}{@{}l}{\textbf{Mean} (18 answerable / 20 for L,\,$t$)}
        & 2.1 & -- & \textbf{290} & \textbf{0.37} & 3.9 & -- & 457 & 2.10 & 2.0 & -- & 973 & 2.62 \\
    \bottomrule
  \end{tabular}

  \vspace{3pt}
  {\scriptsize\raggedright 
  $D$: authoring difficulty on the 1--5 rubric below. 
  $T$: number of attempts needed to reach a correct query ($1$~=~first try). 
  $L$: generated query length in characters. 
  $t$: net database query time in seconds (wall-clock minus fixed client startup).
   $^{\dagger}$ Infeasible from the data (at most one roof material per building exists; no
   roof coverage field); each system answered as a feasibility check. 
   $^{\ddagger}$ 3DCityDB lengths include the shared $\sim$745-char property-pivot CTE prepended to
   every query (i.e., body $+$ CTE). 
   T = 1 everywhere except pykci~C2/E2 (a Cypher
   implicit-grouping fix and an APOC descending-sort fix). \par}

  \vspace{4pt}
  \noindent\textbf{Authoring-difficulty rubric (1--5).}
  \textbf{1}: single attribute + simple aggregate (e.g.\ count, avg).
  \textbf{2}: filter / group-by / \texttt{CASE} value mapping.
  \textbf{3}: multi-step: grid bucketing, cross-tabulation, or a composite score.
  \textbf{4}: requires a mode/argmax (dominant value) or one deep multi-hop traversal.
  \textbf{5}: combines deep traversal(s) \emph{and} aggregation \emph{and} reconstructed
  coordinates in a single query.
\end{table*}

\begin{table*}[h]
    \centering
    \setlength{\tabcolsep}{3pt}
    
    \caption{Per-query performance of locally deployed open-weight LLMs on the \mytoolname\ graph using Ollama.}
    \label{tab:open-llm-detail}
    
    \begin{tabular}{@{}l l cr cr cr @{\hspace{1em}} l l cr cr cr@{}}
    \toprule
    & & \multicolumn{2}{c}{Gemma} & \multicolumn{2}{c}{Mistral} & \multicolumn{2}{c}{Qwen}
    &
    & &
    \multicolumn{2}{c}{Gemma} & \multicolumn{2}{c}{Mistral} & \multicolumn{2}{c}{Qwen} \\
    
    \cmidrule(lr){3-4}\cmidrule(lr){5-6}\cmidrule(lr){7-8}
    \cmidrule(lr){11-12}\cmidrule(lr){13-14}\cmidrule(lr){15-16}
    
    \# & Question & R & $t$ & R & $t$ & R & $t$
    &
    \# & Question & R & $t$ & R & $t$ & R & $t$ \\
    
    \midrule
    
    C1  & City at a glance     & \checkmark & 78  & \checkmark & 64  & \checkmark & 55 &
    E1  & Typology by use type & \checkmark & 98  & \ding{55}  & 63  & \checkmark & 29 \\
    
    C2  & Top roof material \% & \ding{55}  & 36  & \checkmark & 20  & \checkmark & 40 &
    E2  & 1\,km grid + dom.\ mat. & $\circ$ & 101 & \ding{55}  & --  & $\circ$ & 25 \\
    
    C3  & 20 tallest buildings & \checkmark & 25  & \checkmark & 86  & \checkmark & 54 &
    E3  & Solar score 0--7     & \checkmark & 170 & \checkmark & 107 & $\circ$ & 126 \\
    
    C4  & Roof-shape histogram & \checkmark & 22  & \checkmark & 50  & \checkmark & 48 &
    E4  & Storey-height anomaly & \ding{55} & 343 & \ding{55}  & 128 & \ding{55} & 49 \\
    
    C5  & Flat roofs $>5$\,m   & \checkmark & 259 & $\circ$    & 127 & $\circ$    & 50 &
    E5  & Material $\times$ use & \checkmark & 81  & \ding{55} & --  & \checkmark & 58 \\
    
    C6  & Storey histogram     & \checkmark & 25  & \checkmark & 77  & \checkmark & 43 &
    E6  & Dark-roof / 2\,km    & $\circ$    & 349 & \ding{55}  & --  & \checkmark & 41 \\
    
    C7  & Glass roofs          & \checkmark & 148 & \ding{55}  & 114 & \ding{55}  & 59 &
    E7  & Coverage compl.      & \ding{55}  & 184 & $\circ$    & 11  & \checkmark & 30 \\
    
    C8  & High-rises $>50$\,m  & \checkmark & 8   & \checkmark & 42  & \ding{55}  & 35 &
    E8  & High-rise clusters   & \checkmark & 61  & \ding{55}  & --  & \checkmark & 49 \\
    
    C9  & Mixed materials      & \ding{55}  & 17  & $\circ$    & 43  & \ding{55}  & -- &
    E9  & FAR / 1\,km          & \checkmark & 73  & $\circ$    & 342 & \checkmark & 60 \\
    
    C10 & HafenCity bbox       & \ding{55}  & --  & \ding{55}  & 50  & \ding{55}  & -- &
    E10 & Elevation zones      & \checkmark & 58  & \checkmark & 99  & \checkmark & 47 \\
    
    \bottomrule
    \end{tabular}

    \vspace{3pt}
  {\scriptsize
  \raggedright 
  \textbf{R}: result compared to ground truth (\checkmark\ correct, $\circ$ partial, \ding{55} incorrect or failed). 
  \textbf{$t$}: total query time including text-to-Cypher generation and execution (in seconds, local inference). 
  ``--'' indicates a query error with no recorded runtime. 
   Local open-weight LLMs run via Ollama with \texttt{Q4\_K\_M} quantization: Gemma 4 (\texttt{gemma4:31b}, 31.3B), Mistral Medium 5 (\texttt{mistral-medium-3.5:128b}, 128B), and Qwen3.5 (\texttt{qwen3.5:122b}, 125B). 
   Hardware: Intel Core Ultra 9 285K (24 cores, up to 5.7\,GHz), 64\,GB DDR5 ECC RAM, NVIDIA RTX PRO 6000 Blackwell (96\,GB), Linux OS.}
\end{table*}

\end{document}